\documentclass[twocolumns]{aa}
\usepackage{epsfig}
\usepackage{amsmath}
\usepackage{amssymb}
\usepackage{subfigure}
\usepackage{longtable}
\usepackage{graphicx}
\usepackage{color}
\usepackage[breaklinks=true]{hyperref} 
\usepackage{natbib,twoopt}

\usepackage{textcomp}
\usepackage{color}
\usepackage{url}

\usepackage{epsfig}
\usepackage{amsmath}
\usepackage{amssymb}
\usepackage{subfigure}
\usepackage{longtable}
\usepackage{graphicx}
\usepackage{natbib,twoopt}
\usepackage[breaklinks=true]{hyperref} 

\usepackage{textcomp}
\usepackage{color}
\usepackage{url}

\usepackage{graphicx,longtable,lscape}
\usepackage{subfigure}
\usepackage{amssymb}
\usepackage{color} 
\usepackage[normalem]{ulem} 
\RequirePackage[normalem]{ulem} 
\RequirePackage{color}\definecolor{RED}{rgb}{1,0,0}\definecolor{BLUE}{rgb}{0,0,1} 

\begin{document}

\title{Time delays between Fermi LAT and GBM light curves of GRBs}
\author{G. Castignani \inst{1} 
\and D. Guetta\inst{2,3}   \and E. Pian \inst{4,5,6} \and L.  
Amati \inst{4} \and S.  Puccetti \inst{7} \and S. Dichiara \inst{8} }

\institute{
SISSA-ISAS, Via Bonomea 265, I-34136, Trieste, Italy
              \email{castigna@sissa.it}
\and
Osservatorio Astronomico di Roma, via Frascati 33, I-00040 Monteporzio Catone, Italy
\and
Department of Physics and Optical Engineering, ORT Braude, P.O. Box 78, Carmiel, Israel
\and
INAF, Istituto di Astrofisica Spaziale e Fisica Cosmica, Bologna, Via Gobetti 101, I-40129 Bologna, Italy
\and
Scuola Normale Superiore, Piazza dei Cavalieri 7,  I-56122 Pisa, Italy
\and
INFN, Sezione di Pisa, Largo Pontecorvo 3, I-56127 Pisa, Italy
\and
ASI Science Data Center, Frascati, Italy
\and
Dipartimento di Fisica, Universit\`a di Ferrara, via Saragat 1, I-44122, Ferrara, Italy
}

\date{September 9, 2013}

\abstract
{}
{Most gamma-ray bursts (GRBs) detected by the Fermi Gamma-ray Space Telescope  exhibit a delay of up to about 10 seconds between the trigger time of the hard X-ray signal as measured by the Fermi Gamma-ray Burst Monitor (GBM) and the onset of the MeV-GeV counterpart  detected by the Fermi Large Area Telescope (LAT). 
This delay may hint at important physics, whether it is due to the intrinsic variability of the inner engine or related to quantum dispersion effects in the  velocity  of light  propagation from the sources to the observer.  
Therefore, it is critical to have a proper assessment of how these time delays affect the overall  properties of the light curves.}
{We cross-correlated the 5 brightest GRBs of the 1st Fermi LAT Catalog by means of the continuous correlation function (CCF) and of the discrete correlation function (DCF).  The former is suppressed because of the low number counts in the LAT light curves.
A maximum in the DCF suggests there is a time lag between the curves, whose value and uncertainty are estimated through a Gaussian fitting of the DCF profile and  light curve simulation via a Monte Carlo approach.}
{The cross-correlation of the observed LAT and GBM light curves yields time lags that are mostly similar to those reported in the literature, but they are formally consistent with zero.  The cross-correlation  of  the simulated  light curves yields smaller errors on the time lags and more than one time lag for GRBs 090902B and 090926A. For all 5 GRBs, the time lags  are significantly different from zero and consistent with those reported in the literature, when only the secondary maxima are considered for those two GRBs.}
{The DCF method proves the presence of (possibly multiple) time lags between the LAT and GBM light curves in a given GRB and underlines the complexity of their time behavior.  While this suggests that the delays should be ascribed to intrinsic physical mechanisms,  more sensitivity and more statistics are needed to assess whether time lags are  universally present in the early GRB emission and which dynamical time scales they trace.}

\authorrunning {Castignani et al.}
\titlerunning {Time lags in Fermi GRBs}

\maketitle

%

\section{Introduction} 
Gamma ray bursts (GRBs) are the most powerful explosions in 
the Universe.  They have  observed peak luminosities at $\sim$100 keV of
$\sim10^{50}-10^{53}$ erg/s and 
integrated isotropic energy outputs in 10-1000 keV of $\sim10^{51}-10^{54}$ erg, and they are 
detected up to the very early Universe: about a dozen have measured 
redshifts higher than 4 \citep{coward2013}.  A small fraction of GRBs exhibit emission at 
MeV-GeV energies, which was first detected by CGRO-EGRET, more recently by
the AGILE-GRID   \citep[in the 30 MeV-50 GeV energy range,][]{marisaldi2009}, and with more detail and accuracy,  by the 
the Large Area Telescope \citep[LAT,][]{atwood2009} instrument (20 MeV-300 GeV) onboard the Fermi Gamma-ray Space Telescope. 
Possible interpretations have been  given  to explain the paucity of GRBs detected by the LAT  \citep{ghisellini10,guetta11,longo2012}.

The Gamma-ray Burst Monitor \citep[GBM,][]{meegan2009} onboard the Fermi Gamma-ray Space Telescope, operating at energies between 8 keV-40 MeV,  complements the LAT.
The comparison between the Fermi GBM and LAT light curves of GRBs  shows that the onset of the emission of long GRBs
above 100 MeV is systematically delayed by a few  seconds with respect to the start of the GBM signal at hundreds of keV  energies and by a fraction of a second in the case of short and hard GRBs 
\citep[][left panel of their Fig.~2]{abdo2009a,abdo2009b,abdo2009c,giuliani10,delmonte2011,ackermann2011,ackermann2013,piron12}.

That a delay  between the GBM signal onset and the first 
photon detected by LAT is also observed 
in the  brightest LAT GRBs and below 100 MeV, i.e. when photon statistics are relatively rich, suggests that 
this delay is physical and not related to purely 
statistical and instrumental effects.   Moreover, based on
the GBM light 
curve, it is impossible to reproduce the delays using purely statistical methods.
It must be said that the statistical contribution is taken into account in  the 
estimate of the uncertainty of the various temporal parameters, but no correction 
is made on the measurement,  because no plausible high energy emission model would justify
such a correction (R. Bellazzini, private communication).

Two possible physical explanations for this delay have been proposed.  
One invokes different 
emitting regions and mechanisms for the radiation  detected by the GBM and LAT.
It is 
plausible to expect a measurable delay if 
the $\sim$100 keV emission represents the prompt event produced via internal 
shocks \citep{mr99}, and the LAT-detected signal is an aftermath \citep{ghirlanda10}.

The other explanation envisages 
energy-dependent
variation in the speed of light according to quantum gravity (QG) theory \citep{grbgac,nemiroff}.   
It is assumed that the photon momentum is an analytic function of the energy alone.
This can be expanded in a Taylor series, whose linear term is non-zero, to recover the 
classical (non-QG) dispersion relation as the low energy limit. 
Under these assumptions, if we consider a source that produces both high energy  ($E_{high}$)  and
low energy ($E_{low}$) photons the difference $\Delta t$  in the arrival times
between  low and high energy photons is proportional to the ratio between  the
photon energy difference  ($\Delta E=E_{high}-E_{low}$),  and the characteristic QG mass ${\rm M_{QG}}$: $\Delta t= \Delta E/M_{QG} c^2 \times D/c$, where $D$ is the distance of the source and $c$ the speed of light. This idea can be promisingly tested by
accurate arrival-time measurement coupled with a build-up of a
small effect over the huge travel times for the photons from GRBs.
 These time delays have been used to set an upper limit on the Planck mass.
Possible tests of QG in GRBs  have been recently proposed \citep{pavlopoulos2005,gac1,gac2,gac3,vasileiou2013,couturier2013}.

The insufficient accuracy of our
understanding of the physical models,  together with the  Fermi LAT number statistics, make it impossible  to distinguish between these two scenarios without overinterpreting
the data.
However, if either interpretation for the time lags between LAT
and GBM emission is correct (i.e., afterglow vs prompt
emission or QG), we would expect
that the delay affects the entire light curve, not only
the first detected photons.
Before speculating on competing
models that can explain the delays, it is therefore necessary
to ascertain their presence and significance over the
whole GRB evolution in the GBM and LAT energy ranges
\citep[e.g.,][]{delmonte2011}.




In this work we search for delays of the LAT signal with respect to the 
GBM signal in the five brightest LAT GRBs by cross-correlating all of the LAT and GBM light curves.
Our methodology is analogous to that of \citet{ackermann2013c}, who recently applied the DCF to cross-correlating the keV and MeV-GeV light curves of the bright Fermi LAT-detected GRB~130427A.
A similar approach was followed by 
\citet[][]{delmonte2011}. \citet{scargle2008} used a different method assuming a model for the time delay. 

We adopt both the continuous and the discrete correlation function (DCF) methods. The DCF 
was introduced by \citet{ek88} to correlate discrete time series such as
the light curves of active galactic nuclei (AGNs), and it was also applied to GRB light curves 
\citep[e.g.,][]{pian2000}.

Standard correlation function techniques usually require continuous  signals, and therefore  data interpolation, gap filling, and smoothing \citep[e.g.,][]{scargle2010}.
This in turn implies  the suppression of possible rapid variability events,  which are frequent in sources like AGNs and GRBs.  
The primary motive for using the DCF method here is that the observed light curves of GRBs
are discrete and can have different and independent sampling rates.
For regular  and dense time samplings, the DCF yields identical results to a continuous correlation function.
A maximum of the correlation function should indicate a direct correlation of the data trains with a delay at the corresponding time.  However, this could also be spuriously produced or influenced by  statistical fluctuations.
To quantify the significance of each time lag found  using the DCF and to find if our results are affected by the low signal-to-noise (S/N) values we  randomly generate $N\gg1$ GBM 
and LAT light curves according to \cite{peterson1998}.
We correlate them pairwise using the DCF method, and  for each pair of simulated light curves we  estimate the lag at which the DCF peak occurs.  
The distribution of time lags resulting from the N DCFs yields an independent estimate of the time lag itself and of its uncertainty.

The paper is organized as follows. In Sect.~\ref{par:sample} we  introduce the 
GRB sample  and describe the LAT  and GBM data analysis. 
In Sects.~\ref{par:CCFmethod} and \ref{par:DCFmethres} we report our results from application of the CCF and DCF methods, respectively.
In Sect.~\ref{par:DCFsimulations} we report the results of the Monte Carlo 
simulations and in Sect.~\ref{par:disc_concl} discuss our findings.

\section{The sample of the LAT GRBs}\label{par:sample}

We consider the first Fermi-LAT GRB catalog \citep{ackermann2013b} that includes the 
35 GRBs detected by the Fermi LAT instrument from August 2008 to August 2011.
For most of them,  the onset of the LAT 
emission is delayed with respect to the GBM trigger by a few seconds  \citep[or a fraction of a second in the case of short GRBs;][]{piron12}.

To exclude a possible stochastic  origin of this delay related to photon statistics, we selected among these 35 GRBs those with 100~MeV - 10~GeV LAT fluence 
$\geq0.6\times10^{-5}$erg~cm$^{-2}$. This reduces the sample to ten sources. 
Half of them have the test statistic parameter\footnote{The test
statistic parameter is equal to twice the logarithm of the ratio of the
maximum likelihood value produced with a model including the
GRB over the maximum likelihood value of the null hypothesis, i.e., a model that does not include the GRB \citep{ackermann2013b}.} 
TS$<140$, the remaining five have TS$>460$, and four of these have TS$>1450$ \citep[see Table 4 in][]{ackermann2013b}. 
We only retained the sources with TS$>460$, namely 080916C, 090510, 090902B, 090926A, and 110731A. (These also have a boresight angle $\leq 52$~deg.) While GRB090510 has been classified as a short and hard GRB, the others belong to the long duration GRB class.


\subsection{Fermi-GBM data reduction}\label{par:GBM}

The GBM data were retrieved  from the official Fermi site\footnote{\url{ftp://legacy.gsfc.nasa.gov/fermi/data/gbm/bursts}}  and processed with the HEASOFT package (v6.12)  following the Fermi team threads.
We refer to the official Fermi site\footnote{\url{http://fermi.gsfc.nasa.gov/ssc/data/analysis/}}  for details on Fermi data structure and 
analysis.  First,  for each GRB we  considered  the data from both the Bg0 and NaI GBM detectors. 
The Bg0 light curves have a modest S/N, because they are
affected by a high background. Therefore, we rejected  them and we considered  the light curves from the NaI detector alone. 
We further limited our analysis to the 
two NaI detectors  that showed the strongest signal, as inferred from the quick-look 
light curves provided in the Fermi GBM Catalog \citep{goldstein2012}.
The GBM light curves were extracted from the time-tagged event (TTE) files, 
i.e., FITS event files holding information on GRB trigger time and time and energy 
of each photon detected by the corresponding detector.  

We used the {\sc fselect} tool within the software package FTOOLS (HEAsoft 
suite) to filter the event files by selecting  those photons  that have energy between
8~keV and 1000~keV.

The light curves are thus ready for application of the CCF method (Section~\ref{par:CCFmethod}).
To apply the DCF (Sect.~\ref{par:DCFmethres}), the light curves were extracted
and binned into 
time bins of 0.1~s (for the long GRBs in the sample), or 0.01~s (for GRB090510), from 
about 25 s before the trigger time up to 300 s after it, by using the Fermi science 
tool {\sc gtbin}. Then we estimated  the background  at the location of and during each  GRB by fitting  each
light curve  with a first-order polynomial over the 
union of the intervals [-25.0, -20.0]~s and [240.0, 290.0]~s for the long GRBs,
and  over the union of the intervals [-25, -20]~s and [10.0, 200.0]~s for GRB090510. 
(Times are counted from the GBM trigger.) For each GRB, we  subtracted  the 
time-dependent background from the GBM signal of each NaI detector.
For each GRB we summed the two 
background-subtracted GBM light curves, in order to obtain a single GBM light curve 
with a higher S/N.

\subsection{Fermi-LAT data reduction}\label{par:LAT}

For each GRB, we  retrieved LAT Pass 7 Data in a circular region centered  on the 
LAT GRB position, with a radius of ten degrees, and in a time interval spanning 
approximately between -500~s before the GBM trigger and 1000~s after. We processed  the 
data by using the latest version of Fermi Science Tools (i.e., v9r27p1, released in 
April 2012).  We follow the photon selection suggested by the Fermi team for the GRB 
analysis. Therefore, we  used the {\sc gtselect} tool to select only  the events that belong to the ``P7TRANSIENT'' 
or a better class. To maximize photon statistics, they were selected in the full energy range 100 Mev to 300 GeV and with the zenith angle $\le$ 100 deg.
Then we extracted the light 
curves  using the {\sc gtbin} tool and applying a  uniform time bin size 
of 0.1~sec, except for GRB090510, for which we used a bin size of 0.01~sec.

We  subtracted the background from the LAT light curves as follows. For each GRB we 
fitted the light curve  within the union of the intervals [-480.0,-430.0]~s and [940.0,990.0]~s with a 
first-order polynomial. (Times are counted from the GBM trigger.) We subtracted the 
resulting time-dependent background from the LAT signal to obtain the 
background-subtracted light curve. 
The only exception is GRB090926A, for which we prefer to assume a null constant 
LAT background, because this GRB is in the LAT field of view starting only from 30~sec before the GBM trigger.
Furthermore, during the interval [940.0,990.0]~s, Fermi LAT data are rejected from the cleaned file because they have zenith angle >100~degrees. This implies there is no event within the time interval adopted for the Fermi LAT background estimate. We tested that the results outlined in Sects.~\ref{par:CCFmethod} and \ref{par:DCFmethres} do not change if a constant LAT background of 0.1 c/s is adopted.
This value is consistent, as an order of magnitude, with the maximum LAT background estimated for the other GRBs in the sample.

\begin{figure*}[htbp]
 \centering
\subfigure{\includegraphics[width=0.4\textwidth,natwidth=610,natheight=642]{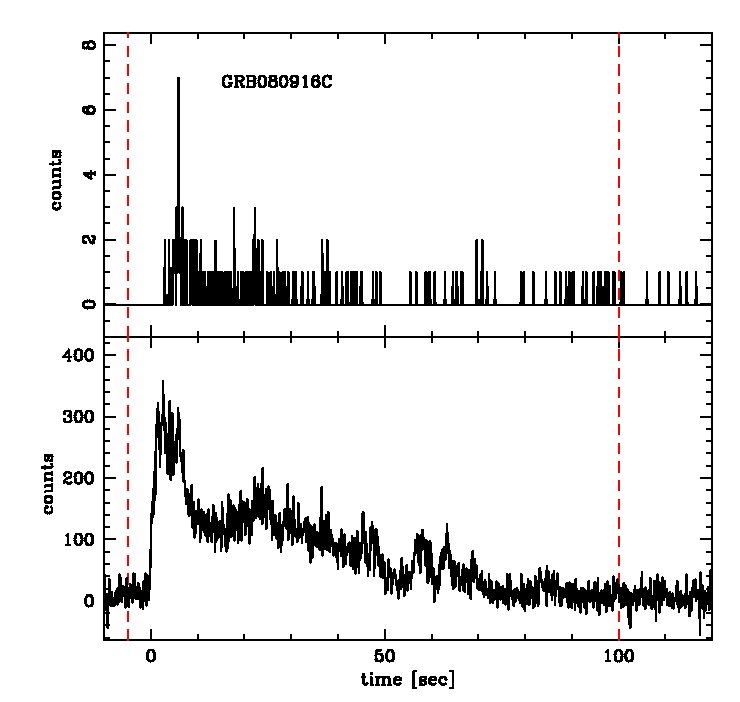}}\qquad
\subfigure{\includegraphics[width=0.4\textwidth,natwidth=610,natheight=642]{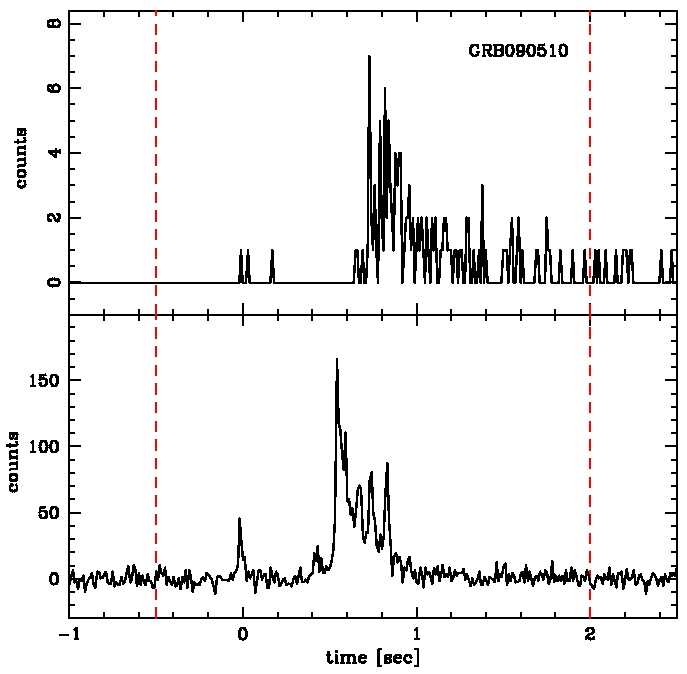}}\qquad
\subfigure{\includegraphics[width=0.4\textwidth,natwidth=610,natheight=642]{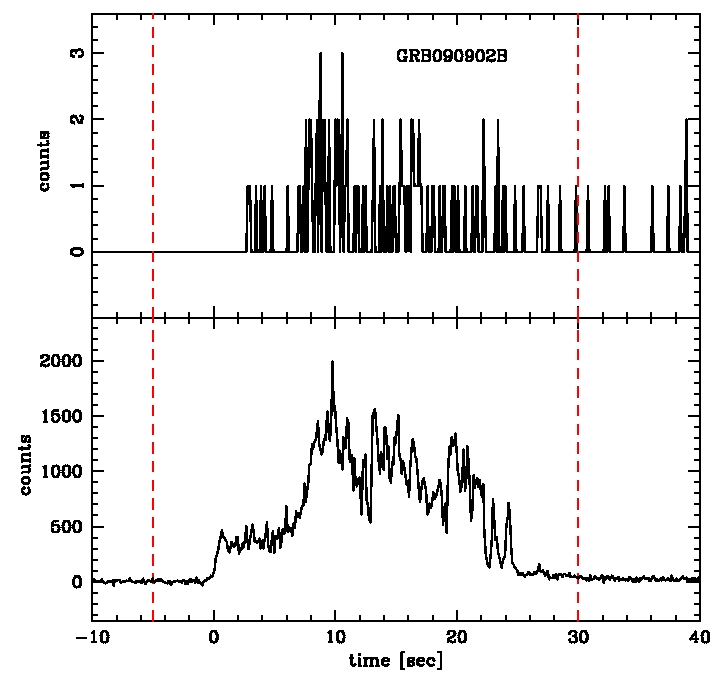}}\qquad
\subfigure{\includegraphics[width=0.4\textwidth,natwidth=610,natheight=642]{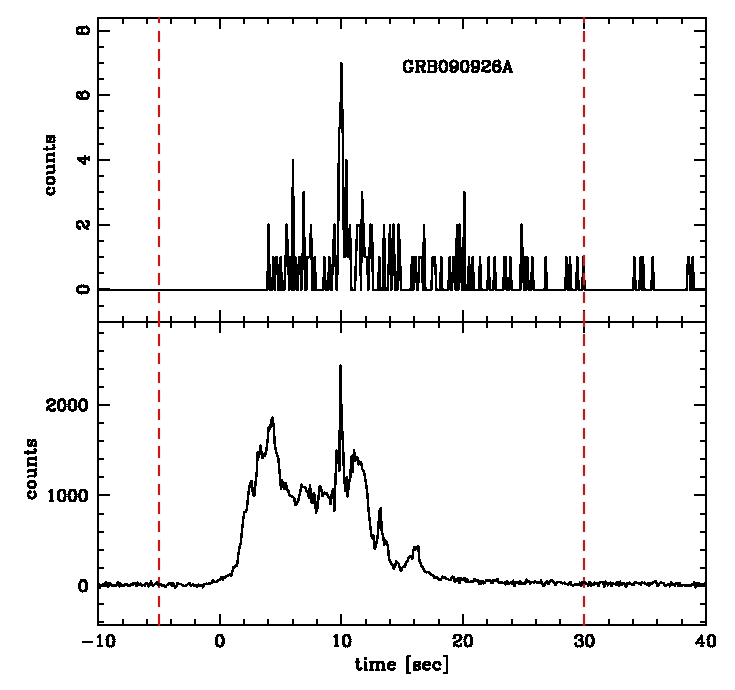}}\qquad
\subfigure{\includegraphics[width=0.4\textwidth,natwidth=610,natheight=642]{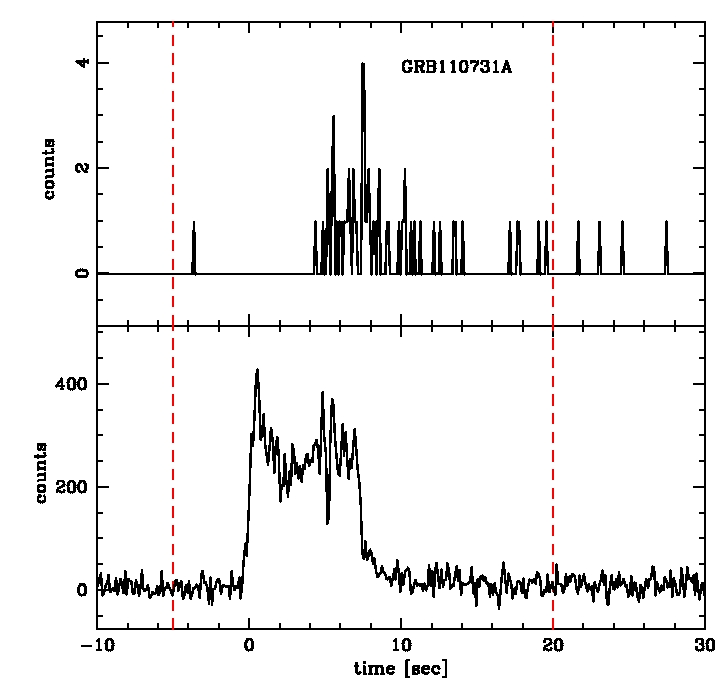}}\qquad
 \caption{Light curves of the 5  GRBs. Top panels: background-subtracted 100 MeV - 300 GeV LAT light curves.  Bottom panels: sum of the background-subtracted light curves from the two NaI detectors with the highest S/N.
Vertical dashed lines show the light curve intervals used for the cross correlation analysis.
}
\label{fig:lc}
\end{figure*}

We note that in the GBM case, such a problem never occurs,
because the GBM raw counts are  always  $\gtrsim$10 per bin.  
In Fig.~\ref{fig:lc} we report the GBM and LAT light curves of   each GRB in our sample.

\section{CCF method and results}\label{par:CCFmethod}
In this section we focus on the continuous correlation function and consider the unbinned GBM and LAT light curves (see Sects.~\ref{par:GBM} and \ref{par:LAT}).  Each curve consists of a time series containing individual photon events. Following the prescription described in \citet{scargle2010}, we performed a 1D Voronoi tessellation of the signals; i.e., we transformed each curve into a succession of rectangles (one for each photon). The basis of each rectangle goes from half way back to the previous photon and ends half way forward
to the subsequent photon. The height is fixed by requiring a unitary area for each rectangle.  The advantage of this representation is that the light curves are now defined at each time and the delta-like discontinuities corresponding to the individual events are reduced to step discontinuities. However, we cannot straightforwardly subtract a background. For this reason, we preferred to consider each GBM NaI light curve separately.  For each GRB, we performed the ordinary correlation function analysis \citep[e.g.,][]{papoulis1965,papoulis1977} between the LAT and the GBM light curves in  their Voronoi representation \citep[see][and references therein]{scargle2010}.
Because of poor number count statistics that mainly affects the LAT light curves, the Voronoi representation assumes low rates at each time. Therefore, the resulting GBM vs. LAT correlation function is similarly suppressed down to null or negligible values. The results of the correlation analysis are therefore inconclusive.

\section{DCF method and results}\label{par:DCFmethres}

The DCF at the lag $\tau$ is estimated by considering all the pairs that are separated in time by an amount between $\tau-\Delta\tau/2$ and
$\tau+\Delta\tau/2$, where $\Delta\tau$ is a specific time binning of the DCF.
Conversely, the CCF at the lag $\tau$ is obtained from all the pairs that are separated in time by exactly 
the amount $\tau$. 

The choice of the time binning for the search of correlation between the two signals is dictated by the need to  sample the possible 
time lag accurately 
(i.e., at least  with $\sim$5 DCF points) and to have decent  S/N 
in the estimate of the DCF in all time bins.  In our case, 
the time lags we are testing are a few seconds ($\sim$0.1~s in the case of GRB090510). Therefore, they are significantly shorter than the duration of the GRB itself 
\citep{abdo2009a,abdo2009b,abdo2009c,ackermann2011,ackermann2013}. 

We subtracted the averages from each of the background-subtracted 
LAT and GBM light curves and applied the DCF method \citep[following ][]{ek88} to these light 
curves over the time intervals marked in  Fig.~\ref{fig:lc}.  
In particular, we weighed the DCF with  the product of the rms dispersions around 
the averages of each of the background-subtracted light curves.
We did not introduce other corrections to the method \citep[][]{white_peterson1994}, such as weighted averages 
or subtraction of individual errors from the light curves variances.
This is  because the noisy signal and the small number counts, especially for the LAT instrument, 
prevented us from estimating 
the uncertainties in the number counts robustly  
and, in turn, the corrections.

\begin{figure*}[htbp]
 \centering
\subfigure{\includegraphics[width=0.39\textwidth,natwidth=610,natheight=642]{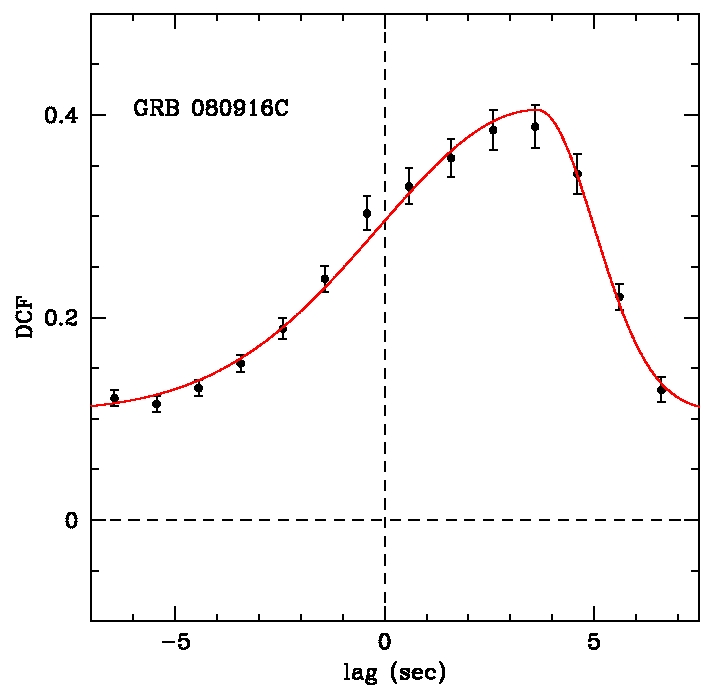}}\qquad
\subfigure{\includegraphics[width=0.39\textwidth,natwidth=610,natheight=642]{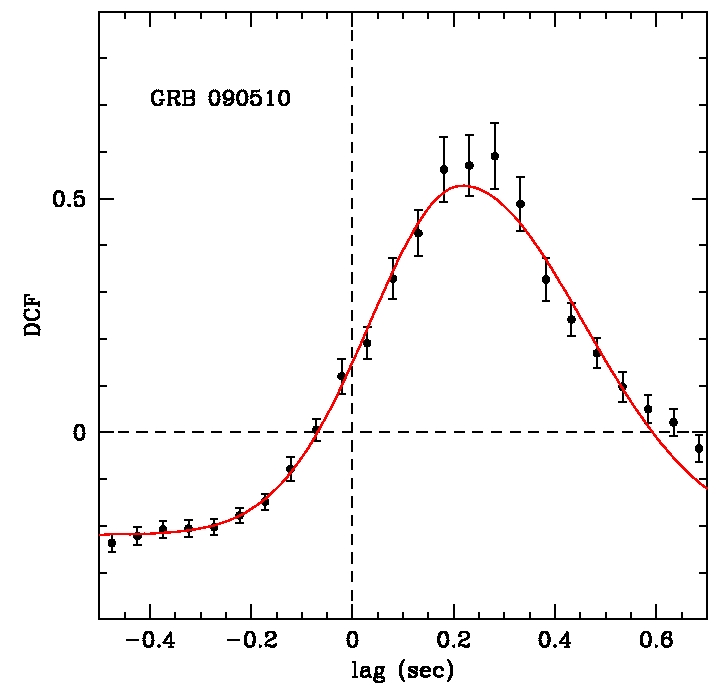}}\qquad
\subfigure{\includegraphics[width=0.39\textwidth,natwidth=610,natheight=642]{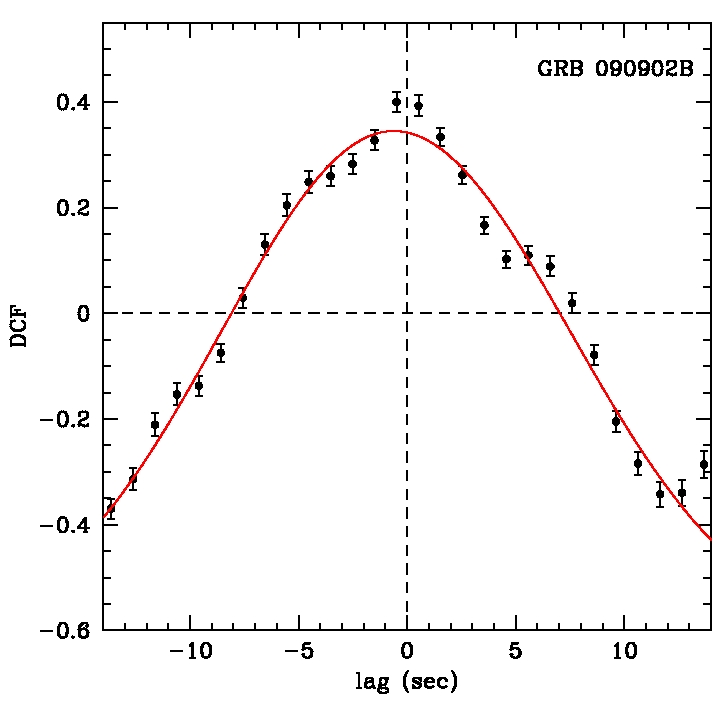}}\qquad
\subfigure{\includegraphics[width=0.39\textwidth,natwidth=610,natheight=642]{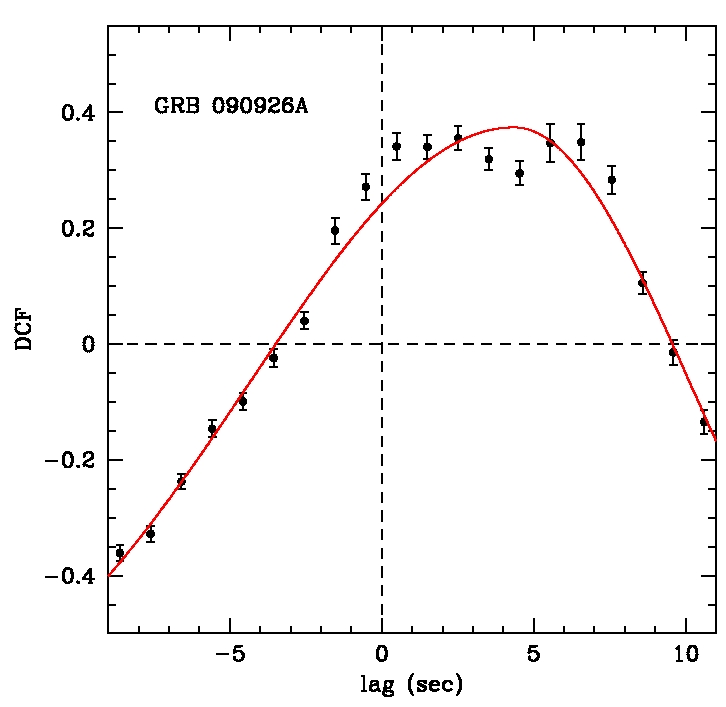}}\qquad
\subfigure{\includegraphics[width=0.39\textwidth,natwidth=610,natheight=642]{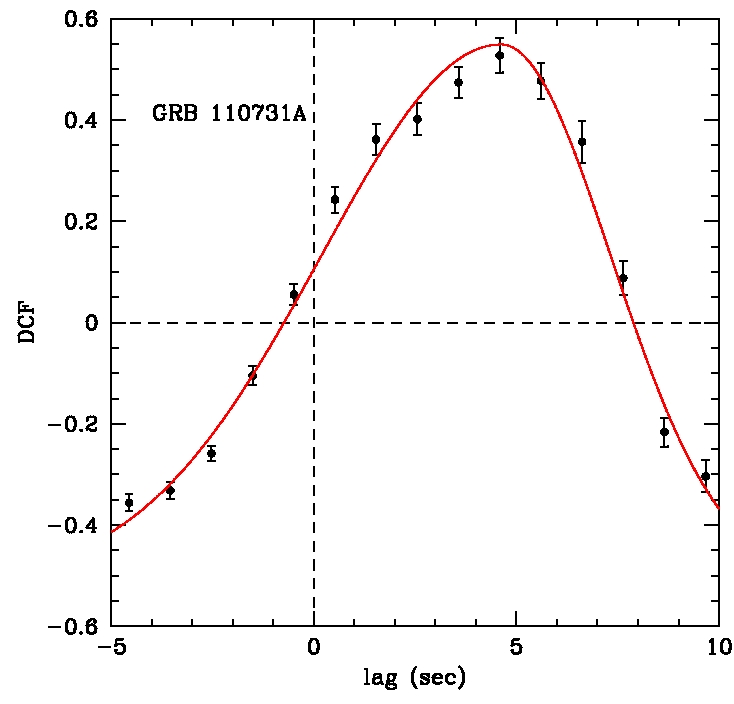}}\qquad
\caption{DCF curves, with individual 1-sigma uncertainties, between the LAT and GBM light curves of our 5 GRBs.  Positive time lags suggest that the LAT is lagging the GBM signal.  The solid red curve is the  asymmetric Gaussian plus constant function that best fits the DCF peak. }
\label{fig:DCF}
\end{figure*}

A DCF time bin of  1~sec was adopted for  the long GRBs and  0.05~sec for GRB090510.
This choice represents an optimal compromise between sufficient number count statistics for each DCF 
bin and  a satisfactory sampling of the time lag between the two signals, as  we verified  a posteriori
by inspecting the DCF. DCFs of our GRBs are reported in Fig.~\ref{fig:DCF}, with their  individual one-sigma uncertainties. 
We   verified that the DCFs do not significantly change if  slightly different DCF time bins 
are adopted. The same holds if  -- for every GRB -- either of the two background-subtracted GBM light curves 
corresponding to the two NaI detectors with the highest S/N, instead of the  sum of them, is considered.  
All DCF curves exhibit a maximum at a formally positive time lag (except for GRB090902B), indicating
that the LAT signal is possibly delayed with respect to the GBM signal.

Since the DCF method does not return any uncertainty on the time lag at which the correlation reaches its maximum, we  estimate the location of the DCF maximum by fitting  the DCF with a constant plus an asymmetric Gaussian function, $f(t)$, over a time interval centered on the peak:

\begin{equation}
 \label{eq:fit}
 f(t)=A+ B\times \begin{cases}
               \exp \left[-\frac{1}{2} \left( \frac{t-\mu}{\Sigma_r}\right)^2 \right]             &{\rm if\hspace{0.2cm}t}~\geq \mu\\
               \hspace{0.05cm}\\
               \exp \left[-\frac{1}{2} \left( \frac{t-\mu}{\Sigma_l}\right)^2 \right]             & {\rm otherwise}
           \end{cases}
.\end{equation}

This approach is similar to that of \cite{zhang99}, although a  symmetric Gaussian satisfied their purposes.  
It  is a simplified and empirical assumption:  there is no theoretical and observational 
reason to prefer one specific form for the function $f$.  
The Gaussian assumption provides both a rough estimate for the uncertainty 
on the time lag and a robust estimate of the time lag itself. They are given by 
the Gaussian dispersion and the lag at which the $f$-function reaches its maximum ($\mu$), respectively.
We also estimate the formal error associated with the $\mu$ parameter according to the $\chi^2$ statistics.
However, the $\Sigma_r$ and $\Sigma_l$ parameters are better estimators for the
uncertainty associated with the time lag than the formal error associated with the $\mu$ parameter.
This is because the $\Sigma_r$ and $\Sigma_l$ parameters represent the extent of the DCF spreading around its maximum.
In fact, the width of the DCF maximum shows the range of time lags within which the correlation peak occurs. 
Conversely, the $\mu$ parameter error, estimated with the $\chi^2$ statistics, represents how well our model 
fits the data. Therefore, such an estimate is clearly model dependent and physically unrelated to the 
actual time lag uncertainty.


In Table~\ref{table:fit} (Cols. 2 to 6) we report the best-fit parameters, along with the corresponding one-sigma formal errors 
obtained by the $\chi^2$ parameter marginalization \citep{cash1976}. 
In Fig.~\ref{fig:DCF} we plot the best-fit curves. 


All DCFs exhibit clear peaks of correlation, while anti-correlation is excluded; i.e., no negative minimum is observed.  
The time lags of the correlations are positive, with the exception of GRB090926A, but they are never significantly 
different from zero at the 1-$\sigma$ level (see Cols. 3, 4, 5 in Table \ref{table:fit}).

\begin{table*}
\caption{Best-fit parameters for the time lags between observed and simulated LAT and GBM light curves.}
\label{table:fit}
\centering
\begin{tabular}{cccccccccc}
\hline\hline
GRB  &  A$^a$ &  $\mu^a$ (s)  &  ${\Sigma_r}^a$ (s) & ${\Sigma_l}^a$ (s) &  B$^a$ &  lag$^b$ (s)   &  lag$^c$ (s)   & T$^d$ (s) & Ref.$^e$ \\
 (1)  & (2)  &  (3)  &  (4)  &  (5)  &  (6)  &  (7)  &  (8)  &  (9)  &  (10)  \\
\hline
080916C & 0.11$\pm$0.01 & 3.65$\pm$0.22 &  1.36$\pm$0.17  &  3.81$\pm$0.22       & 0.30$\pm$0.01   & $3.79^{+0.26}_{-0.55}$ &  --  & $\sim$3.6 & 1 \\

090510 & -0.22$\pm$0.02 & 0.22$\pm$0.02  & 0.24$\pm$0.02  & 0.18$\pm$0.02   & 0.75$\pm$0.02   & $0.21^{+0.03}_{-0.06}$  & --  & $\sim$0.56 & 2 \\

090902B & -0.62$\pm$0.09 & -0.62$\pm$0.21 & 8.15$\pm$0.67  &   7.96$\pm$0.59    & 0.97$\pm$0.09  & $-1.52^{+2.01}_{-1.00}$ & $6.18^{+1.31}_{-0.80}$ & $\sim$9.6 & 3 \\

090926A & -0.75$\pm$0.09 & 4.36$\pm$0.26  & 5.78$\pm$0.36 &  8.70$\pm$0.72  & 1.12$\pm$0.09    & $2.32^{+0.87}_{-0.94}$ & $5.52^{+0.32}_{-0.64}$  & $\sim$3.3 &  4 \\

110731A &  -0.47$\pm$0.03 & 4.62$\pm$0.17 & 2.71$\pm$0.16 & 4.45$\pm$0.24  & 0.98$\pm$0.03   & $4.07^{+0.23}_{-0.18}$ & --  &  $\sim$2.4 & 5 \\
\hline
\multicolumn{9}{l}{$^a$ Best fit parameters with formal 1-$\sigma$ errors for the  constant plus  asymmetric Gaussian function (Eq.~\ref{eq:fit}). The}\\
\multicolumn{9}{l}{ ~~   1-$\sigma$ asymmetric uncertainty on the $\mu$ parameter is given by the $\Sigma_r$ and $\Sigma_l$ parameters.} \\ 
\multicolumn{9}{l}{$^b$ Time lag and  1-$\sigma$ uncertainty derived from the  maximum of the time lag distribution of simulated light curves.}  \\
\multicolumn{9}{l}{$^c$ Time lag and  1-$\sigma$ uncertainty derived from the secondary maximum of the time lag distribution of simulated }\\
\multicolumn{9}{l}{ ~~ light curves.}  \\
\multicolumn{9}{l}{$^d$ Time lags based on the initial LAT delay, as reported in the literature.} \\
\multicolumn{9}{l}{$^e$ References for Col.~9:  1.~\cite{abdo2009a};   2.~\cite{abdo2009b}; 3.~\cite{abdo2009c}; }  \\
\multicolumn{9}{l}{ ~~ 4.~\cite{ackermann2011}; 5.~\cite{ackermann2013}.}  \\

\end{tabular}
\end{table*}

\section{Monte Carlo simulations and results}\label{par:DCFsimulations}

As stressed by \cite{zhang99},  a  constant plus  Gaussian function,
although representative of the peak position and the dispersion for the  DCF, does not necessarily provide a 
statistically adequate fit.  Furthermore, the presence of peaks in the correlation might be due to statistical fluctuations more than real time delays between the two signals.
For these reasons we  estimated
 the significance and the uncertainty of each  time lag by means of Monte Carlo simulations, 
as prescribed in \cite{peterson1998}.  

We apply the so-called ``flux randomization'' (FR) method \citep{peterson1998} to simulate $N$ 
light curves starting from the  
GBM and LAT observed, not background-subtracted,  light curves. 
According to the FR procedure, we start from a certain light curve and produce a number $N$ of simulated signals that are drawn from a specific distribution (depending on the nature of the signal) whose average is equal to the observed light curve value, for each time.
We applied the FR procedure to both the GBM and LAT light curves for each GRB.
The simulated light curves are then correlated pairwise by means of the DCF, as described in Sect.~\ref{par:DCFmethres}. This provides a GBM vs. LAT time lag distribution that is used to 
derive the actual delay (if any is present) of the LAT emission with respect to the GBM emission.

We chose $N = 10,000$ for each GRB in our sample.
Given an observed light curve $\{F_0, F_1, ...,  F_j,... \},$ we randomized the j-th number count at the time $t_j$ 
assuming a  Poisson distribution around the average $\phi_j$, which is set equal to the flux $F_j$, if it  is positive.  Otherwise, we set $\phi_j$ equal to the background level estimated at the time $t_j$ for the given GRB and  the considered detector. 
For GRB090926A  we were not able to estimate an LAT background (see Sect.~\ref{par:LAT}).  Therefore, for this GRB alone, where $F_j = 0$, we assume $\phi_j=\beta\times\Delta$t, 
where $\Delta$t is the bin of the LAT light curve, and $\beta$ is chosen to be 0.01~c/s and 0.1 c/s. These values are consistent with the LAT background estimated for the other GRBs in the sample. 
Then, for all GRBs, we  subtract the  background from each simulated light curve.
Assuming Poisson uncertainties for the observed number counts 
only leads to small deviations from the more 
correct estimate  that is generally used for small number counts, as is the case for LAT light curves
\citep[i.e., N$<$10, see, e.g.,][]{gehrels86}.

For each GRB, the 2$\times N$ GBM NaI simulated light curves are pairwise
summed in such a way that the two simulated curves in a given pair never come 
from the same NaI detector light curve  and that each of the 2$\times N$ NaI simulated light curves belongs to one and only one pair.

Then, we have $N$ simulated LAT and $N$ simulated GBM background-subtracted light curves,
 which we compared pairwise with the DCF method, as in Sect.~\ref{par:DCFmethres}. 
We chose  DCF time bins equal to those adopted for the correlation of the observed light curves: 1.0~sec for  the long GRBs, 0.05~sec  for the short GRB090510.  As in Sect.~\ref{par:DCFmethres},
we limited the DCF analysis to the time intervals of the GBM and LAT light curves designated in  Fig.~\ref{fig:lc}. 

We fit the  $N$ DCFs  obtained by cross-correlating the simulated GBM and LAT  light curves 
with  a constant plus asymmetric Gaussian function  ($f(t)$, see Eq.~\ref{eq:fit}).
The fit is performed with the minimum-$\chi^2$ method.
This is analogous to what was done in Sect.~\ref{par:DCFmethres} for the DCF obtained by
cross-correlating the observed light curves.  

\begin{figure*}[htbp]
 \centering
\subfigure{\includegraphics[width=0.38\textwidth,natwidth=610,natheight=642]{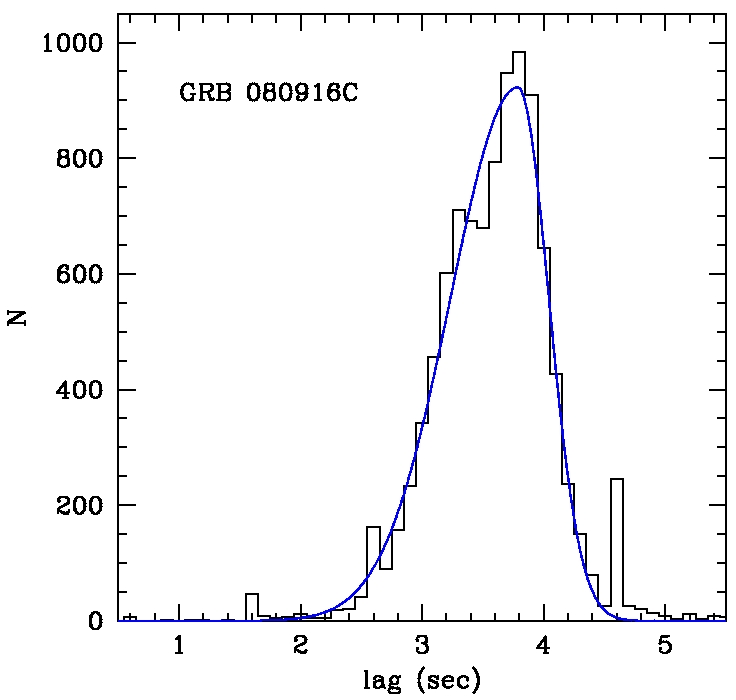}}\qquad
\subfigure{\includegraphics[width=0.38\textwidth,natwidth=610,natheight=642]{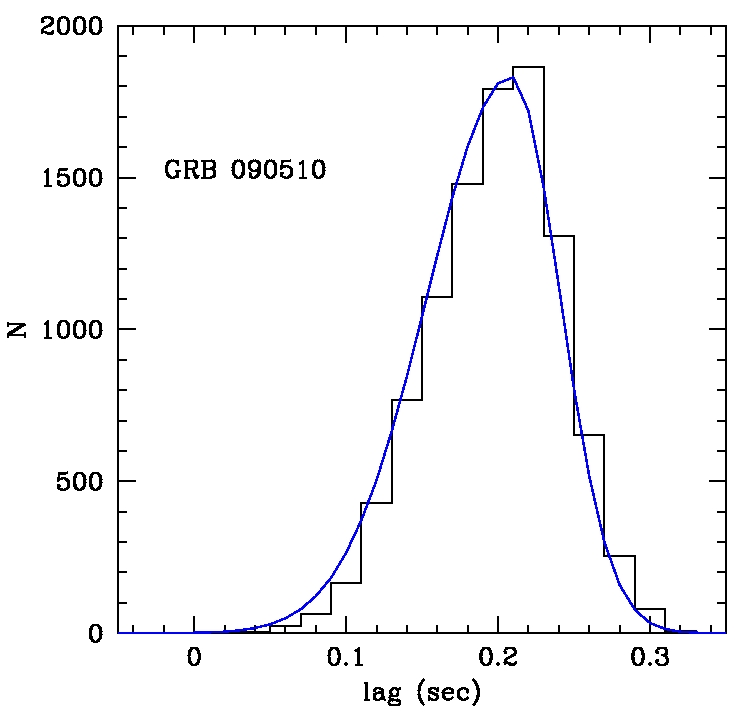}}\qquad
\subfigure{\includegraphics[width=0.38\textwidth,natwidth=610,natheight=642]{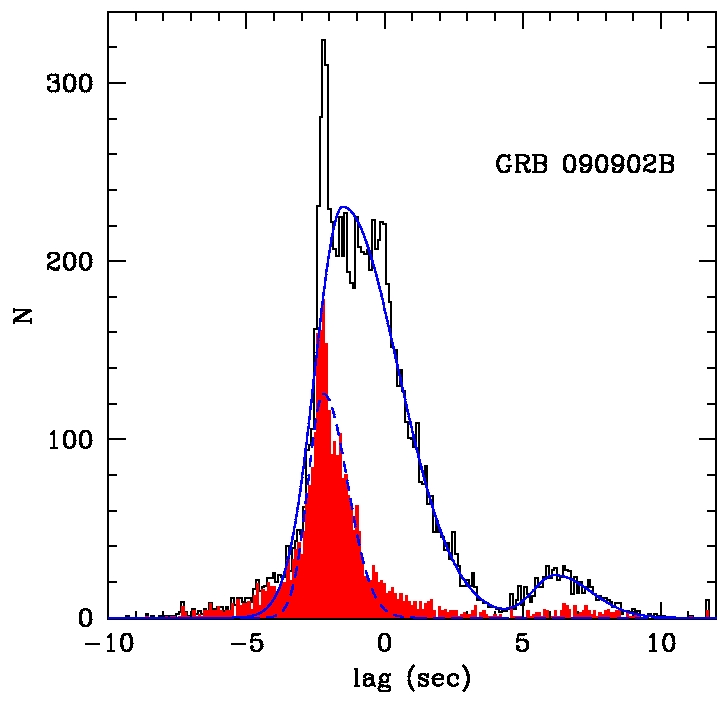}}\qquad
\subfigure{\includegraphics[width=0.38\textwidth,natwidth=610,natheight=642]{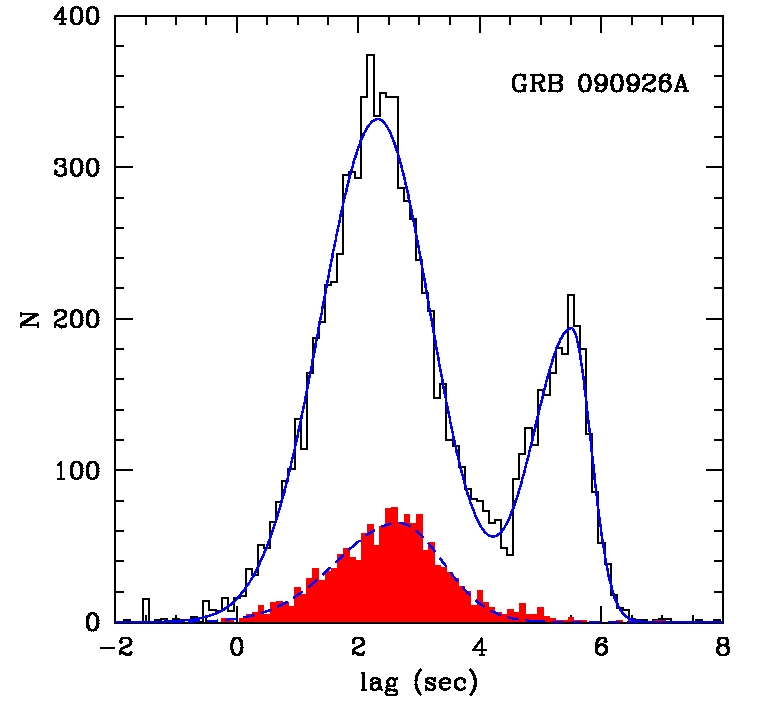}}\qquad
\subfigure{\includegraphics[width=0.38\textwidth,natwidth=610,natheight=642]{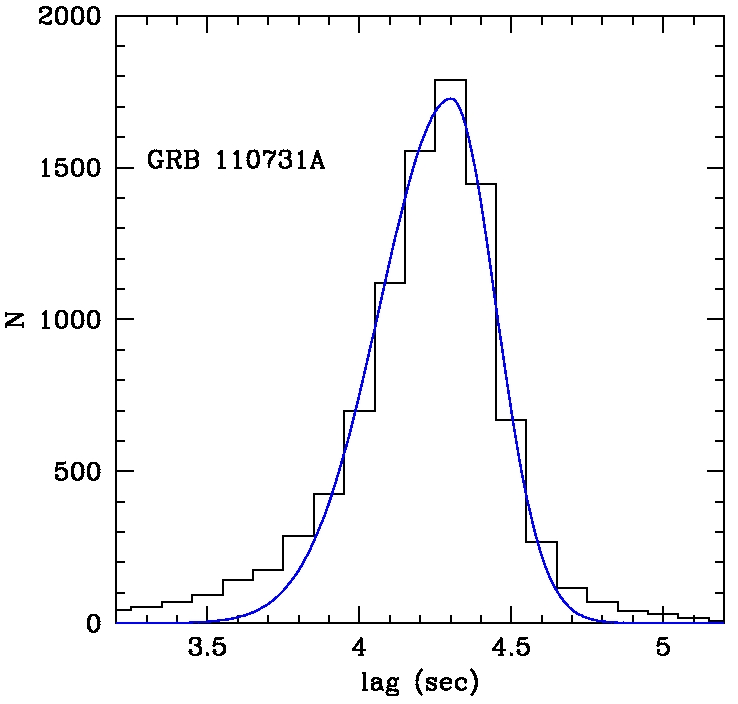}}\qquad
 \caption{Time lag distributions (black histograms)  obtained  by cross-correlating the simulated GBM and LAT light curves.  Positive time lags correspond to the LAT signal lagging the GBM signal.   
The asymmetric Gaussian fit is reported as a blue solid line.  The red areas correspond to the distributions of time lags obtained
considering those fits that are significant at a $4\sigma$ level or higher, according to the $\chi^2$ statistics 
(i.e., fits with $\chi^2/{\rm dof}\gtrsim$3 are rejected).
The asymmetric Gaussian fit for the red areas is reported as a blue dashed line.
}
\label{fig:lag_distribution}
\end{figure*}

In Fig.~\ref{fig:lag_distribution} we report the resulting distributions of best-fit time lags.
These distributions are fitted with an asymmetric  
Gaussian\footnote{For $N\gg1$, it is reasonable to expect that the distribution of the best-fit time lags
is a Gaussian, because of the central limit theorem. Therefore, 
the role of the asymmetric Gaussian is completely different here from, and independent
of, the asymmetric Gaussian used to fit the DCF curves.}, since they are evidently asymmetric on either side of the maximum.   
In Table~\ref{table:fit}  (Col.~7) for each GRB we report the corresponding best fit parameters
 (i.e., the lag at which the Gaussian reaches its maximum; the reported uncertainties are the square roots of the two Gaussian variances).   
Concerning GRB090926A,  we plot only the results corresponding to $\beta=$0.1~c/s, since we have verified that they are similar for $\beta = $0.01~c/s.

The time lags that are obtained according to this procedure are consistent 
with those derived from the observed DCFs (Col.~3). With the exception of GRB090902B,  they are also 
compatible with those reported by the LAT team based on the initial delay of the LAT with respect to the GBM signal (Col.~9).
 However, since the uncertainties are smaller than those determined from the Gaussian fits of observed DCFs 
in Sect.~\ref{par:DCFmethres}, the time lags of GRBs 080916C, 090510A, and 110731A 
 are significantly different from zero at the 3-$\sigma$ level.   

In the case of GRBs~090902B and 090926A, the time lag distributions  suggest that there are secondary maxima
at $\sim$6~s (see Fig.~\ref{fig:lag_distribution} and Table~\ref{table:fit}, Col.~8). 
Close inspection of Fig.~\ref{fig:DCF} shows that the presence of two maxima 
is also marginally seen in the DCF of the observed light curves.
This is because  a modest deviation 
from a clear single Gaussian fit occurs for both of these DCF: 
a shoulder (at about $\sim5$~s) and two bumps (at $\sim$2.5 s and $\sim$6 s) are present in the DCF of GRB090902B and GRB090926A,
respectively.

 For both GRBs, the secondary maxima  are consistent with 
the time lags reported by the LAT team (Table~\ref{table:fit}, Col.~9). 
However, by excluding those DCF fits that are significant at  a $4\sigma$ level or lower, according
to the $\chi^2$ statistics (i.e., by rejecting those fits with $\chi^2{\rm /dof}\gtrsim$3),
these secondary maxima become much less significant (see Fig.~\ref{fig:lag_distribution}).

\section{Conclusions}\label{par:disc_concl}
Motivated by the detection of time delays between the onset of LAT and GBM signals in Fermi-detected GRBs, 
we have adopted the DCF method to cross-correlate the LAT and GBM light curves of the five brightest LAT GRBs  in the first Fermi LAT GRB catalog
and thus to estimate the delay between the arrival times of MeV-GeV and $\sim$100-keV energy photons over the whole 
time evolution of the GRBs.  
We searched for delays both in the observed light curves and in 
light curves that were randomly generated via Monte Carlo simulations.

From the DCF of the observed light curves, we derived the time lags using a constant 
plus an asymmetric Gaussian approximation of the DCF maximum and determined
the formal errors associated with a Gaussian fit (Table~\ref{table:fit}, Cols.~2-6).
The reliability of these uncertainties depends on the correctness of the Gaussian approximation of the DCF profile 
around its maximum.

For each GRB, we also performed the  individual DCFs  of simulated light curves and
 estimated the associated time lags, analogously to
what was done for the DCF obtained by using the observed light curves.
This  allowed us to independently estimate both the time lag and its uncertainty 
as the average and the square root of the variance of the Gaussian fit function, respectively (Table~\ref{table:fit},  Cols.~7 and 8).
The estimates obtained by adopting the time lag distributions are more robust than those obtained by using 
the individual DCF that results from the observed light curves. 
This is because the estimates derived from the individual DCFs are based on a statistical approximation, rather than on a functional form description of the individual DCFs.

The time lags derived from  the DCFs of the observed light curves  are all formally consistent with zero, 
although they are mostly similar to those reported in the literature.
This result is analogous to what is reported by \cite{delmonte2011}, who, using the cross-correlation approach,  do not recover statistical significance for the  time delay of $\sim$10 s observed between the start time of the  AGILE MCAL and GRID light curves of GRB100724A.

When the simulated light curves are cross-correlated and the resulting time lag distributions are  fitted with 
Gaussian functions,  in three cases (i.e. GRBs 080916C, 090510A, 110731A) the best-fit time lags are significantly different from zero and compatible with those
reported in the literature.    For GRB090902B, the main time lag ($\sim$~-2~s)  is formally very different from what has been previously reported \citep[9.6 s,][]{abdo2009b}, but not significantly different from zero. 
For  GRB090926A, the main time lag (2.3 s) is  consistent with the one reported by \cite{ackermann2011}, but again not significantly different from zero.    However, GRBs 090902B and 090926A also have secondary maxima in their time lag distributions.
For both GRBs they  correspond to time lags that are significantly different from zero and similar to those previously 
reported in the literature.
We note that the secondary maxima become less significant when only the satisfactory fits ($\chi^2{\rm /dof}\lesssim3$)
 are retained.  The reason may be that, since the fits corresponding to the secondary maxima generally
have a more limited significance than those associated with the primary maxima, 
the DCF curves where they are more prominent have a complex morphology and are not well fit. 

The presence of these secondary maxima suggests a complexity in the time behavior of the gamma-ray signals.
While in general our results suggest that  the cross-correlations are influenced by the observed initial delays between the
LAT and GBM light curves, they also  show other time scales and  suggest  that these delayed LAT signal onsets are probably due to 
intrinsic physics.  A systematic cross-correlation analysis on a bigger sample than used here may 
set better constraints on the physical origin of the time delays.

\begin{acknowledgements}
We  thank T. Alexander, R. Bellazzini, L. Bildsten, N. Omodei, and E. Waxman for stimulating  discussions.
DG and EP are grateful for hospitality 
at the Weizmann Institute of Science in Rehovot, Israel, where part of this work was developed.
This work was partially supported by INAF PRIN 2011 and ASI/INAF contracts I/009/10/0 and I/088/06/0. 

\end{acknowledgements}


\begin{thebibliography}{50}
\bibitem[Abdo et al.(2009a)]{abdo2009a} Abdo, A. A., et al., 2009a, Science, 323, 1688 
\bibitem[Abdo et al.(2009c)]{abdo2009c} Abdo, A. A., et al., 2009b, Nature, 462, 331  
\bibitem[Abdo et al.(2009b)]{abdo2009b} Abdo, A. A., et al., 2009c, ApJ, 706, 138  
\bibitem[Ackermann et al.(2011)]{ackermann2011} Ackermann, M., et al. 2011, ApJ, 729, 114 
\bibitem[Ackermann et al.(2013a)]{ackermann2013} Ackermann, M., et al. 2013a, ApJ, 763, 71 
\bibitem[Ackermann et al.(2013b)]{ackermann2013b} Ackermann, M., et al. 2013b, ApJS, 209, 11
\bibitem[Ackermann et al.(2013c)]{ackermann2013c} Ackermann, M., et al. 2013c, arXiv:1311.5623
\bibitem[Amelino-Camelia et al.(1998)]{grbgac} Amelino-Camelia, G., Ellis, J., Mavromatos, N.~E., Nanopoulos, D.~V., \& Sarkar, S., 1998, Nature, 393, 763
\bibitem[Amelino-Camelia et al.(2013a)]{gac1} Amelino-Camelia, G., Fiore, F., Guetta, D., \& Puccetti S., 2013a, submitted to PRX, arXiv:1305.2626
\bibitem[Amelino-Camelia et al.(2013b)]{gac2} Amelino-Camelia, G., Guetta, D., \& Piran T. 2013b, submitted to JCAP, arXiv:1303.1826
\bibitem[Atwood et al.(2009)]{atwood2009} Atwood, W. B., Abdo, A. A., Ackermann, M., et al., 2009, ApJ, 697, 1071
\bibitem[Cash(1976)]{cash1976} Cash, W.\ 1976, \aap, 52, 307
\bibitem[Couturier et al.(2013)]{couturier2013} Couturier, C., Vasileiou, V., Jacholkowska, A., Piron, F., et al., 2013, arXiv1308.6403
\bibitem[Coward et al.(2013)]{coward2013} Coward, D. M., Howell, E. J., Branchesi, M., et al., 2013, MNRAS, 432, 2141
\bibitem[Del Monte et al.(2011)]{delmonte2011} Del Monte, E., Barbiellini, G., Donnarumma, I., et al., 2011, A\&A, 535, 120
\bibitem[Edelson \& Krolik(1988)]{ek88} Edelson, R. A., \& Krolik, J. H. 1988, ApJ, 333, 646
\bibitem[Gehrels(1986)]{gehrels86} Gehrels, N., 1986, ApJ, 303, 336
\bibitem[Ghirlanda et al.(2010)]{ghirlanda10} Ghirlanda, G., Ghisellini G., \& Nava, L., 2010, A\&A, 510, 7
\bibitem[Ghisellini et al.(2010)]{ghisellini10} Ghisellini, G., Ghirlanda, G., Nava, L., \& Celotti, A. 2010, MNRAS, 403, 926
\bibitem[Giuliani et al.(2010)]{giuliani10} Giuliani, A., Fuschino, F., Vianello, G., et al., 2010, ApJ, 708, 84
\bibitem[Goldstein et al.(2012)]{goldstein2012} Goldstein, A., Burgess, J. M., Preece, R. D., et al., 2012, ApJS, 199, 19
\bibitem[Gonzalez et al.(2003)]{gonzalez03} Gonzalez, M. M., Dingus, B. L., Kaneko, Y., et al., 2003, Nature, 424, 749
\bibitem[Guetta et al.(2011)]{guetta11} Guetta, D., Pian, E., \& Waxman, E., 2011, A\&A, 525, 53
\bibitem[Guetta(2013)]{gac3} Guetta, D., 2013, arXiv:1303.1619, Invited Review, 2012 Fermi Symposium proceedings - eConf C121028
\bibitem[Longo et al.(2012)]{longo2012} Longo, F., et al., 2012, A\&A, 547, 95 
\bibitem[Marisaldi et al.(2009)]{marisaldi2009} Marisaldi, M., G. Barbiellini, E., Costa, et al. 2009(astro-ph/0906.1446)
\bibitem[Meegan et al.(2009)]{meegan2009} Meegan, C., et al. 2009, ApJ, 702, 791
\bibitem[Meszaros \& Rees(1999)]{mr99} Meszaros, P., \& Rees, M. J., 1999, MNRAS 306, pp. 39-43
\bibitem[Nemiroff et al.(2011)]{nemiroff} Nemiroff, R. J., 2011, AIPC, 1358, 83
\bibitem[Papoulis(1965)]{papoulis1965} Papoulis, A., 1965, Probability, Random Variables, and Stochastic Processes (McGraw-Hill: New York)
\bibitem[Papoulis(1977)]{papoulis1977} Papoulis, A., 1977, Signal Analysis, (McGraw-Hill: New York)
\bibitem[Pavlopoulos(2005)]{pavlopoulos2005} Pavlopoulos, T. G., 2005, PhLB, 625, 13
\bibitem[Peterson et al.(1998)]{peterson1998} Peterson, B. M., Wanders, I., Horne, K., et al., 1998, PASP, 110, 660
\bibitem[Pian et al.(2000)]{pian2000} Pian, E., Amati, L., Antonelli, L. A., et al., 2000, ApJ, 536, 778
\bibitem[Piron et al.(2012)]{piron12} Piron, F., et al., 2012, Proceedings of the Gamma-Ray Bursts 2012 Conference(GRB 2012). May 7-11, 2012. Munich, Germany
\bibitem[Press et al.(1992)]{press92} Press, W., et al. 1992, Numerical Recipes: The Art of Scientific Computing(2d ed.; Cambridge: Cambridge Univ. Press)
\bibitem[Scargle et al.(2008)]{scargle2008} Scargle, J. D., Norris, J. P., \& Bonnell, J. T., 2008, ApJ, 673, 972
\bibitem[Scargle(2010)]{scargle2010} Scargle, J.~D., arXiv1006.4643
\bibitem[Takeshi et al(2010)]{takeshi2010} Takeshi, U., et al., 2010, American Astronomical Society, HEAD meeting 11, 2.05; Bulletin of the American Astronomical Society, Vol. 41, p.654
\bibitem[Vasileiou et al.(2013)]{vasileiou2013} Vasileiou, V., Jacholkowska, A., Piron, F., et al., 2013, PhRvD, 87, 122001
\bibitem[White \& Peterson(1994)]{white_peterson1994} White, R. J., Peterson, B. M., 1994, PASP, 106, 879
\bibitem[Zhang et al.(1999)]{zhang99} Zhang, Y. H., et al., 1999, ApJ 527, 719
\end{thebibliography}
\end{document}